\documentclass{article}

\usepackage{arxiv}

\usepackage[utf8]{inputenc} 
\usepackage[T1]{fontenc}    
\usepackage{hyperref}       
\usepackage{url}            
\usepackage{booktabs}       
\usepackage{amsfonts}       
\usepackage{nicefrac}       
\usepackage{microtype}      
\usepackage{lipsum}
\usepackage[pdftex]{graphicx}
\usepackage{textcomp} 

\title{Student and Teacher Meet in a Shared Virtual Reality: A one-on-one Tutoring System for Anatomy Education}

\author{
  Patrick Saalfeld \\
  Department of Simulation and Graphics \\
  Otto-von-Guericke University Magdeburg \\
  Germany \\
  \texttt{saalfeld@isg.cs.uni-magdeburg.de} \\
  \And
  Anna Schmeier \\
  Department of Simulation and Graphics \\
  Otto-von-Guericke University Magdeburg \\
  Germany \\
  \And
  Wolfgang D'Hanis \\
  Institute of Anatomy \\
  Otto-von-Guericke University Magdeburg \\
  Germany \\
  \And
  Hermann-Josef Rothkötter \\
  Institute of Anatomy \\
  Otto-von-Guericke University Magdeburg \\
  Germany \\
  \And
  Bernhard Preim \\
  Department of Simulation and Graphics \\
  Otto-von-Guericke University Magdeburg \\
  Germany \\
}

\begin{document}
\maketitle

\begin{abstract}
We introduce a Virtual Reality (VR) one-on-one tutoring system to support anatomy education.
A student uses a fully immersive VR headset to explore the anatomy of the base of the human skull. A teacher guides the student by using the semi-immersive zSpace. Both systems are connected via network and each action is synchronized between both systems.

The teacher is provided with various features to direct the student through the immersive learning experience.
She can influence the student's navigation or provide annotations on the fly and, hereby, improve the students learning experience.
The system is implemented using the \textit{Unity} game engine.
A qualitative user study demonstrates that the one-on-one tutoring approach is feasible and sets a solid base for future research in the area of shared virtual environments for anatomy education.
\end{abstract}


\section{Introduction}
\label{introduction}

An in-depth understanding of anatomical structures is fundamental for medical diagnosis and treatment.
There exist a variety of  methods to teach anatomy, comprising the use of anatomy atlases and dissections.
These classical methods are complemented by computer-based systems that provide interactive 3D visualizations, motivating tasks and feedback \cite{Preim:2018}.
The rise of computer-based anatomy education systems is also due to a shortage of cadavers as well as more severe constraints of time and resources, such as available teaching staff \cite{SAK10}.
Recently, immersive VR systems have been developed and evaluated to further improve anatomy education.
The immersion is promising to enable the imagination of complex anatomical regions with a high density of structures that even may be intertwined.

This work is motivated by the limitations of self-guided learning.
Although existing VR applications enable students to freely explore a virtual environment at an individual pace, they may waste time following unproductive paths or may miss important details \cite{DHB01}.
We consider the \emph{guidance} of a student through a VR experience as promising to ensure effective learning.
In a \textit{one-on-one tutoring} scenario, a teacher could give instructions on finding a specific structure and intervene if a student takes long in navigating.
In this paper, we introduce an approach for a VR one-on-one tutoring system to provide a guided approach to VR-based anatomy education.
Focusing on \textit{gross anatomy}, the human base of the skull will serve as an application example.

VR devices, such as head-mounted displays (HMD), provide position tracking and 3D input devices, enabling students to interact naturally in the virtual world.
Students can inspect virtual models of the human body and understand spatial relations through freely adjusting the view. 
They benefit from a direct integration of names and descriptions of specific structures, linked to the virtual models -- a fully immersive 3D anatomy atlas.
There are first applications that use this potential for anatomy education \cite{Marks:2017,Maresky:2019,Pohlandt:2019}.

For the teacher, however, an HMD is not a suitable device, as the user's real environment is completely blocked.
This makes it more difficult for the teacher to access prepared notes, use a keyboard for text input  and look out for the student.
Therefore, we implemented the teacher system on a semi-immersive device, i.e. a display that provides stereo rendering by wearing 3D glasses and motion parallax by tracking the user's head.
For this, we chose the zSpace.
While semi-immersive systems lack the power of full immersion, as they constrain the stereopsis to the frontal visual field, they have advantages that make them the better choice for certain applications \cite{WAB93}, e.g. allow the user to still be aware of her surroundings.

The presented system, developed in the Master thesis by Schmeier~\cite{Schmeier2017}, supports a medical student in exploring anatomical structures with different navigation techniques intuitively.
Additionally, dedicated input techniques are provided to the teacher system to guide the student to a learning session by creating annotations and help the student navigate to specific locations.
The two systems are connected via network whereas the state of the student, the teacher and the learning environment are synchronized.

\section{Related Work}
In this section, we discuss research relevant for a VR tutoring approach in anatomy. 

\noindent\textbf{Computer-Assisted Anatomy Education.}
Human anatomy is a key aspect in medical education.
Medical curricula target the understanding of anatomical structures, their form, position, size and their spatial relationships, e.g. which nerve supplies a particular muscle.
The scientific study of anatomy covers different aspects, such as \textit{gross anatomy} (the study of the structure and positioning of organs on a macroscopic level), \textit{histology} (the study of microscopic anatomy), and \textit{neuroanatomy} (the study of the brain, spinal cord and peripheral nervous system) \cite{BHBSPFD07}.
\textit{Gross anatomy} is the focus of this work.

The \textit{VOXEL-MAN} \cite{HPPRSST96} pioneered three-dimensional anatomy teaching.
It provides a detailed high-quality 3D model of the human anatomy with hundreds of segmented anatomical structures that can be inspected from arbitrary angles.
Virtual dissection can be performed on the 3D virtual anatomy.
The spatial information (shape, size, texture and subdivisions) is linked to the symbolic domain (verbal description, anatomic concepts and relationships) by means of labelled and textual descriptions.

Another early example is the \textit{Zoom Illustrator} \cite{PRS97}, a zoomable user interface where 3D models of the anatomy are adapted to the visibility of labels and textual descriptions.
The \textit{VOXEL-MAN} and the \textit{Zoom Illustrator} are inspired by the atlas metaphor, i.e., high quality illustrations and their textual descriptions in an anatomy atlas were used for orientation.
The work of Kraima et al.~\cite{kraima2013toward} focuses on one region to create a comprehensive anatomical atlas of the pelvic region.

Alternatively, a 3D puzzle metaphor motivates stronger to actually employ 3D interaction facilities.
The gaming component of a puzzle is a further benefit for learning.
Ritter et al. introduced a 3D puzzle for anatomy education providing various techniques to support depth and shape perception \cite{RPDS00}.
This idea was extended by Pohlandt et al.~\cite{Pohlandt:2019} and used in a VR environment.3D anatomy education benefits from web-based systems. While early systems required the user to install plugins, such as VRML, with the advent of WebGL, interactive 3D visualization was available in web browsers without the need to install any additional software. Birr et al.~\cite{Birr:2013} created one of the first such systems for studying liver anatomy.
In the survey of Preim and Saalfeld, a wide variety of anatomy education systems is discussed~\cite{Preim:2018}.

\noindent\textbf{VR as a Learning Environment.}
In 1990, Bricken \cite{Bri90} pointed out that the characteristics of VR are the same as those of good teaching.
A teacher wants to create an environment that is programmable, meaning a curriculum as a planned sequence of instruction, which the students participate in.
VR augments learning with experience and replaces the desktop metaphor with a world metaphor \cite{Pso95}.
Regian et al.~\cite{RSM92} state that VR interfaces are more motivating than traditional 2D interfaces.
There is a certain excitement over the use of new technologies \cite{Pso95}.
This can be used to make learning more interesting and fun, which may make students remain engaged in learning for a longer period of time.
The immersion of VR systems is seen as a strong benefit and educators from different fields become increasingly interested in taking advantage of VR technology \cite{HS08}.

VR technology has been integrated in educational applications \cite{HRL10}.
Students benefit from the natural way in which they can interact.
As the cognitive effort is reduced compared to non-immersive systems, users can fully focus on the scenario rather than on the semantics of the interface \cite{Pso95, HV97}.

Usually, VR systems for education are based on free-choice learning and discovery \cite{TN02}.
Following a museum metaphor, users are often encouraged to stroll around the virtual world and inspect details further, according to personal interest.
This way, they will remember general things and random details, instead of global concepts and overarching principles \cite{TN02}.
Important learning material may be missed, and moreover, it can be time-consuming when the user follows unproductive paths \cite{DHB01}.
To ensure that all important facts are covered, at least a partly prescribed course is needed.
A way to achieve this is to let the teacher engage in the virtual environment and guide the student to ensure efficient learning.

\noindent\textbf{Shared VR.}
Immersive VR applications usually provide a one-person experience.
However, letting people share a VR experience has huge potential.
Brown~\cite{Bro00} outlines the potential of shared VR for education and states, that network performance, good graphics and low latency are required.
VR enables new opportunities to bring together students and teachers from remote places.
It can provide access to education everywhere, circumventing problems that the isolation of distance learning usually brings \cite{BPT01} such as following unproductive learning paths.
It raises new possibilities for explanation and guidance, as educators can lurk over the shoulders of students and intervene \cite{Pso95}. 

Previous studies show that a collaboration between multiple learners in a shared immersive virtual environment can have a positive educational effect \cite{JTW99, MJOG99, JF00}.
The \textit{NICE Project} \cite{JRLVBM98} is an early example of a collaborative immersive VR learning environment.
It is targeted towards children and uses the \textit{CAVE}.
The remote users are represented with avatars that mimic the movement of the user, comprising head movement and gestures.
This supports the sense of presence and immersion.
However, the influence of a teacher being present in a shared virtual environment is mainly unevaluated \cite{TN02}, even though it is likely to have a positive effect on learning.

Different scenarios for a shared VR application can be classified, according to the relationship between the users.
In a \textit{symmetric setting}, each user is an instance with equal conditions.
They have the same capabilities when interacting with the system.
Opposed to this, an \textit{asymmetric setting} provides each user with different interaction possibilities.
Through unique perspectives, each user contributes to the virtual world in his individual way.
Most collaboration examples of multiple students interacting in one shared virtual environment are symmetric approaches.

An example for an asymmetric setting can be found by Le Chénéchal et al. \cite{LDLGRA16}.
They designed 3D manipulation tasks collaboratively, while the users have a different view on the scene and different manipulation possibilities.
One user has an observer view point.
Using the semi-immersive \textit{zSpace}, she can perform manipulation tasks from a distance.
A second user is placed inside the object that is manipulated.
Having a detailed view, she can perform precise interactions wearing a VR headset and motion controller.
This approach can be transferred to an education scenario: a teacher may have the global view to supervise a student, while the student has a more detailed view through an immersive device.

\noindent\textbf{One-on-One Tutoring in VR.}
Tutoring describes a one-on-one dialogue between a teacher and a student with the purpose of helping the student to learn something \cite{EM06}.
Bloom~\cite{Blo84} showed that tutoring is a very powerful tool of instruction.
Compared to other methods such as group instruction, tutoring is a much more effective way of learning.
Despite the great learning success through one-on-one tutoring, it is not always applicable, due to high costs or the number of needed instructors.
Computers can support one-on-one tutoring, delivering learning material appropriate for individual needs at "any time, any place, any pace" \cite{DHB01}.
Intelligent tutoring systems, where an AI takes over the part of the tutor, are a research area in itself.
Natural language dialogue is very important to tutoring \cite{EM06}.
Even though intelligent tutoring systems try to cope with this, one-on-one tutoring by humans is still worthwhile.

\section{A One-on-One Tutoring Approach for Shared VR}
This section is divided into the two roles teacher and student in order to identify the specific requirements for both sides of the system.
The fully immersive system for the student is introduced 
followed by the semi-immersive VR system for the teacher.

\subsection{The Education Scenario}
The main objective is to improve a medical student's learning experience.
The free exploration method should be enhanced by allowing a teacher to guide the student in a shared virtual environment.
The teacher should be able to help the student to navigate through the virtual world in order to find  structures, as well as give explanations using annotations.
The teacher also supports the student if questions arise.

The scenario considers a one-on-one setting involving a teacher and one student.
The student explores a virtual anatomy model through a fully immersive VR headset.
As the student's real environment is completely blocked out, due to the worn HMD, the teacher needs to join the virtual environment in order to guide the student.
To enable such a shared virtual environment, a network connection is necessary to synchronize information.
Both teacher and student will be located in the same physical room while using the system.

\subsubsection{Medical Background on the Human Skull Base}
The focus of this work is on gross anatomy and the human skull base.
We interviewed an anatomist with 12 years of experience in teaching anatomy to obtain background information on the skull base as well as which aspects are the most important to learn.
Firstly, learning the anatomy of the skull is an integral part of anatomy curriculum.
Textbooks about human anatomy~\cite{Cun18, Ham82} describe the human skull in detail and anatomy atlases \cite{PP00} contain illustrations from different perspectives to convey spatial relations.
Illustrations in anatomy atlases use a consistent color scheme to distinguish arteries (red), veins (blue) and nerves (yellow). These color schemes should be used in the virtual learning environment.
The term \textit{skull} (cranium) is used to describe the entire skeleton of the head.
The \textit{base} of the skull describes the lower area of the skull.

The base of the human skull has several holes (foramina) through which arteries, veins and cranial nerves (those that directly emerge from the brain) pass.
One objective of medical students is to learn these foramina of the human base of the skull by their name and location, together with the structures that pass them.
Hereby, the precise course of the structures is secondary.
Students primarily have to understand which organs they further lead to.
Regarding the cranial nerves, it is essential to understand, what their specific tasks are (i.e. control muscles or glands or transmit sensory impressions).

\subsection{Requirements for Different Roles: Student and Teacher}
\label{sec:requirements}
Having two distinct sides of the system, with users that follow different goals, a separate analysis of their specific requirements is necessary.
Understanding the roles that the student and teacher occupy, helps to decide, which devices and input techniques are suitable.
In our scenario, the student uses a VR headset and the teacher a semi-immersive display.
This approach corresponds with an asymmetric setting, similar to the work of Le Chénéchal et al.~\cite{LDLGRA16}, where asymmetric object manipulation is described.

\subsubsection{Student System}
The target group for the student system is an undergraduate student, who is enrolled in a medical curriculum.
Here, anatomy education is an integral part in the first years of study \cite{SAK10}.
The student's main goal is to learn about the anatomy of the human skull base to acquire knowledge about foramina and their passing structures. This leads to the following requirements:

\noindent\textbf{Immersion.}
Fully immersive VR interfaces are considered to provide a great benefit for learning as they enhance motivation and the feeling of immersion augments learning with experience~\cite{Preim:2018}.
Thus, a fully immersive VR headset is an appropriate choice for the student system.

\noindent\textbf{Information is Directly Linked to Structures.}
To enhance learning, the student needs to develop a cognitive model of the anatomical structures.
This is facilitated when the spatial and symbolic domain of information are linked appropriately. Therefore, 3D models should be augmented with their names and descriptions.

\noindent\textbf{Recognize Small Details.}
In an evaluation of interaction techniques for the exploration of 3D illustrations \cite{PPS99} it was observed that medical students tend to enlarge the virtual model strongly to recognize small details. Since the VR environment is not bound to space limitations such as the real world, an enlargement of the anatomic structures should be possible.

\noindent\textbf{Natural and Intuitive Navigation without Disorientation.}
The student needs to be able to navigate through the virtual environment with minimal cognitive effort and without disorientation.
Position tracking and motion controllers allow a natural and intuitive interface for this.

\noindent\textbf{Inspection of Single Structures.}
Single structures, such as nerves, may be partly occluded by other structures.
If the student wants to be able to inspect a structure closely, selection and manipulation techniques need to be implemented.

\subsubsection{Teacher System}
The main goal for the educator using this system is to support a medical student in the acquisition of knowledge about the anatomical structures of the human skull base.
Based on the student system, the teacher system needs to provide an interface to intervene in the virtual environment that is explored by the student.
This leads to the following requirements for the teacher system:

\noindent\textbf{Balance of Immersion and Access to the Real World.}
As the teacher is familiar with the anatomy of the human base of the skull, the benefits of full immersion are less important.
Instead, the teacher might need to have access to prepared notes or a keyboard for text input. Considering this, a fully immersive VR system that completely blocks out the virtual world is not appropriate for the teacher's needs. This makes a semi-immersive system the best fit.

\noindent\textbf{Awareness of Student's Position and Orientation.}
In order to perform guidance, the teacher has to be aware of the student's location within the virtual environment at any time.
Thus, the student needs to be represented at the according location on the teacher's system.

\noindent\textbf{Support the Student's Navigation.}
The teacher should help the student navigate through the virtual environment.
In order to lead her to a specific location, a possibility to hint the student needs to be realized.
Further, facing the problem that self-guided exploration can take an unnecessarily long time, the teacher needs to be able to completely alter the position of the student and place her at a desired location.

\noindent\textbf{Provide Annotations for the Student.}
The creation of content for the symbolic domain has do be provided at run time.
Textual input is best suited for annotations of names and descriptions.
The spatial and symbolic domain can be linked effectively with labels.
If the teacher wants to highlight a certain area or illustrate concepts, a free-form 3D sketching tool is appropriate. 

\noindent\textbf{Change of the View.}
To perform the above-mentioned tasks an intuitive way of navigation needs to be implemented for the semi-immersive system which allows the teacher to focus on a desired location.

\subsection{The Shared Virtual Environment}
Since the student needs to recognize small anatomy details, an oversized skull model is placed in the virtual environment.
The student being represented much smaller, can navigate within the model while obtaining a micro view of the base of the skull.
She is immersed in an environment where the bones of the skull form the walls around her.
Thus, small details can be explored without the need of further scaling.

\begin{figure}[htb]
	\centering
	\includegraphics[width=\linewidth]{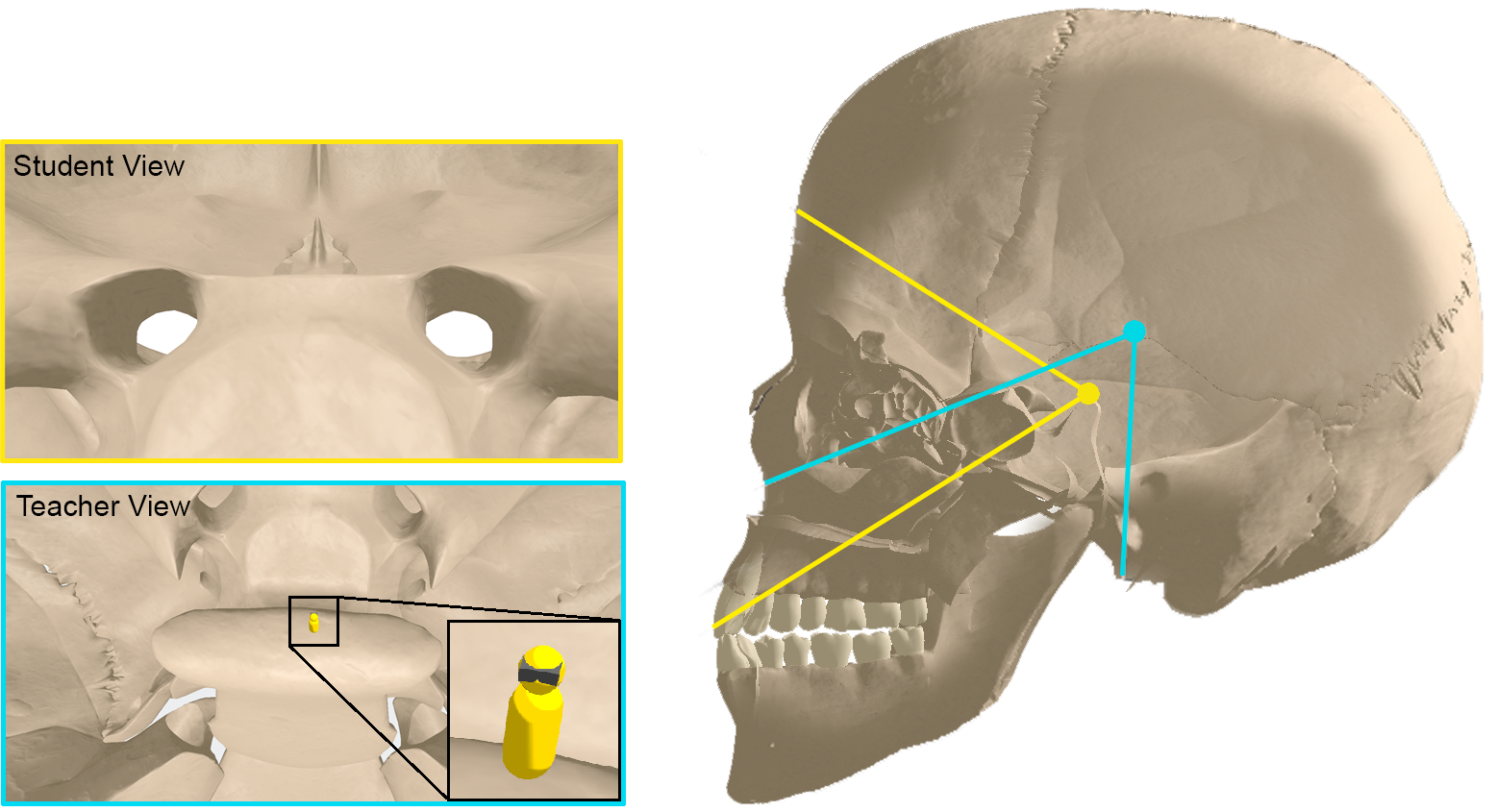}
	\caption{Initial position, orientation and view of student and teacher within the virtual environment. The teacher view shows an avatar of the student.}
	\label{fig:environment}
\end{figure}

The teacher, on the other hand, takes an \textit{overview position} within the environment.
She looks down at the student, in order to follow her position and orientation.
Figure~\ref{fig:environment} illustrates this asymmetric setting and shows the initial position, orientation and the corresponding view of both users. 

The student is represented as a simplified, cel-shaded~\cite{LC00} avatar on the teacher system in order to contrast with the environment (see Fig.~\ref{fig:environment}).
This way, the teacher can track the student's position and orientation.
The teacher, on the other hand, is not represented in the student system.

\subsection{Fully Immersive Student System}
There exist several possibilities to embed a student in an immersive environment.
HMDs are a favorable choice, as they are easier to integrate within a learning environment.
They are becoming more affordable and the setup is less problematic than the setup of systems such as a \textit{CAVE}.
Even though, mobile HMDs are cheaper and easier to use with several students at a time, six degrees of freedom (6 DOF) input devices and movement is not possible and their hardware does not allow demanding 3D applications.
The \textit{HTC Vive} is chosen due to its room-scale position tracking capabilities and two 6 DOF controllers.
The following sections present the user interface of the student system and discuss its functionality in detail.

\subsubsection{User Interface}
The student has one HTC Vive controller in each hand.
The left controller is the \textit{Menu Controller}.
It displays a menu, while the right controller as \textit{Interaction Controller} enables selection through ray-casting and is used for navigation.
Both controllers can be swapped to suit left-handed and right-handed users equally.

\begin{figure}[htb]
	\centering
	\includegraphics[width=\linewidth]{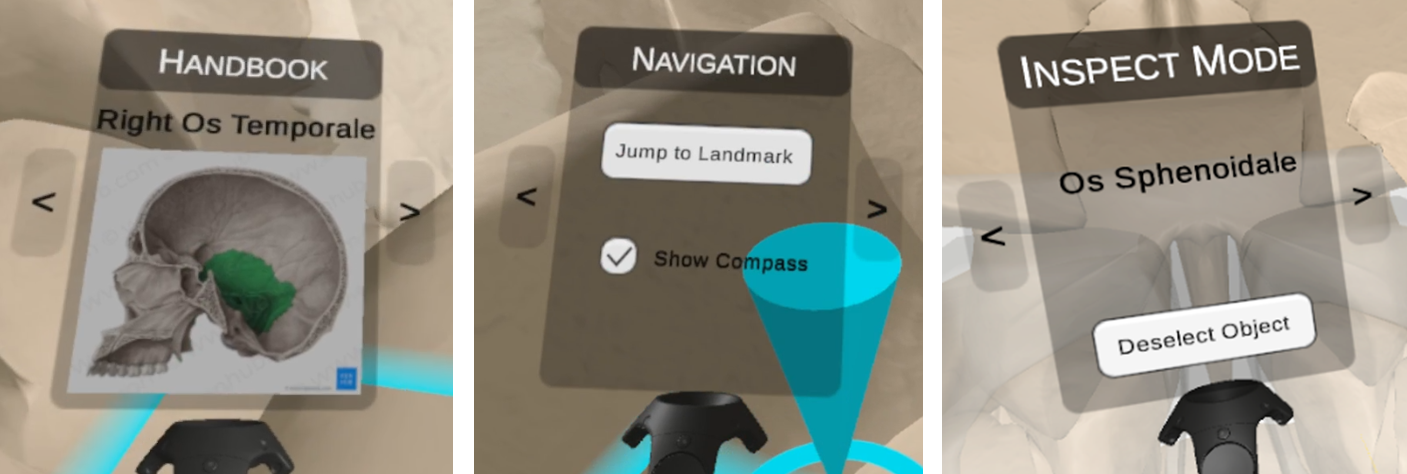}
	\caption{The three menu pages of the student system, where each represents a different interaction mode: left -- handbook, middle -- navigation, right -- inspect.}
	\label{fig:vive_menu}
\end{figure}

\noindent\textbf{Menu Controller.}
A menu is attached on top of the controller.
It is part of the 3D scene and moves with it.
This placement should create a \textit{handbook metaphor}.
This way, the student can adjust the readability of textual content by bringing the menu closer to her eyes.
Information about the virtual objects within the environment are displayed here and the user can browse through different menu pages to change the interaction mode.
Figure \ref{fig:vive_menu} shows the three menu pages.
The corresponding modes and the according functionalities are outlined in the following.
The pages can be browsed by using the left and right side of the touchpad, respectively.

\noindent\textbf{Interaction Controller.}
The \textit{Interaction Controller} should be used with the dominant hand as it is used for precision tasks such as pointing and selecting.
A laser beam originates from the controller and follows its pointing direction.
This way, distant objects can be pointed at and selected without the need of navigation towards them.
The first selectable object that the ray hits is highlighted. Additionally, the controller's vibro-tactile feedback is used to strengthen the sense of having hit something.
The trigger on the back of the controller is used to select the object.

\subsubsection{The Handbook Functionality}
The handbook is the default mode when the system is started and assists the student in exploring the virtual world.
Whenever the ray hits a sub-structure of the human skull base, its name, a short description or a 2D illustration is shown (see Fig.~\ref{fig:vive_menu}, left).

\subsubsection{Navigation Methods}
The student system offers three different ways for navigation, i.e. walking, free flying, and teleportation, which are described in the following.

\noindent\textbf{Walking.}
As the \textit{HTC Vive} offers a room-scale experience with position tracking, physical walking can be mapped to the virtual environment.
The available space for walking is indicated through a barrier in the virtual world to prevent a collision with walls.
This intuitive and natural navigation can be used for maneuvering around in a limited area to get different viewing angels on a location of interest. 

\noindent\textbf{Free Flying.}
As the virtual environment is larger than the available space for walking in the real world, another way of navigation needs to be offered to the student.
Corresponding to a \textit{flying carpet} metaphor the student can fly freely in every direction.
The barrier that indicates the space for walking serves as orientation and gives the feeling of standing on a virtual platform.

The flying is controlled with the touchpad of the interaction controller.
During pressing the touchpad, the student flies in the direction of the \textit{Interaction Controller}.

\textit{Free Flying} can be used at any time and is not dependent on the active mode.
The movement speed is predetermined to a fast walking speed.
This is a compromise between moving relatively quickly and keeping the orientation inside the skull. 
If larger distances need to be  overcome, teleportation can be used.

\noindent\textbf{Teleport.}
The teleportation navigation mode ca be accessed over a dedicated menu page (see Fig. \ref{fig:vive_menu}, center).
When activated, this mode offers the student either to directly jump to a landmark that was defined by the teacher or teleport to an area that is chosen by the student via pointing and selecting.

\begin{figure}[htb]
	\centering
	\includegraphics[width=0.215\linewidth]{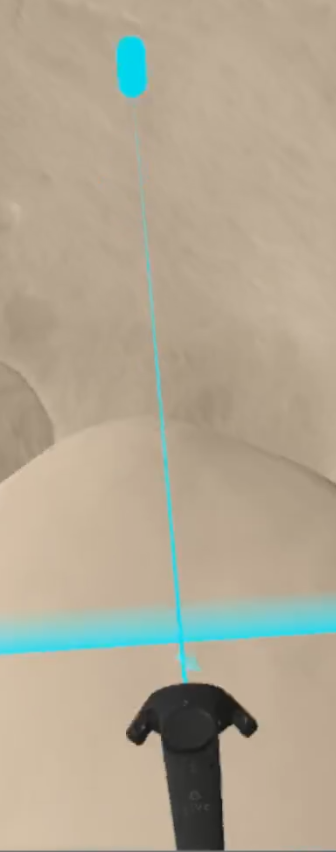}
	\includegraphics[width=0.745\linewidth]{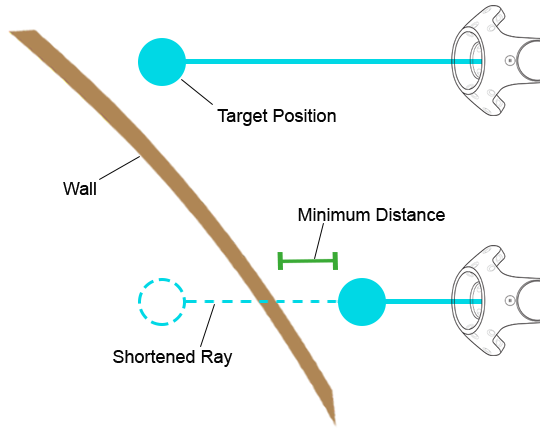}
	\caption{Left: the adjustable ray for teleporting shows the target location through a capsule at the tip. Right: in order to avoid teleporting through a wall, the teleportation ray is shortened until there is a minimum distance between the wall and the target location.}
	\label{fig:teleport_concept}
\end{figure}

When the \textit{Navigation Mode} is active, the standard yellow-orange interaction laser beam is replaced with a cyan teleport laser beam. 
The length of this beam can be adjusted by swiping up or down on the touchpad of the interaction controller.
At the tip of the beam, a capsule is shown that represents the target position after teleporting (see Fig. \ref{fig:teleport_concept}).
Even though this is a fast method to overcome longer distances, the student may lose orientation.
This could be weakened by transitioning between the old and new position with a continuous motion. This, however, can lead to \textit{cybersickness}. Therefore, the teleportation is realized discretely.

The student should only be able to teleport to locations in sight in order to reduce the danger of lost orientation.
Thus, teleporting through walls or within a wall should not be possible.
Figure \ref{fig:teleport_concept} illustrates how we prevented this.
If the student would teleport through the wall, the length of the laser beam is shortened automatically to a distance that assures a minimum distance to the wall.
This, however, also prevents students from teleport into small corridors.
If this is desired, she is still able to do so with the free flying technique.

\subsubsection{Inspecting Single Structures}
\begin{figure}[htb]
	\centering
	\includegraphics[width=\linewidth]{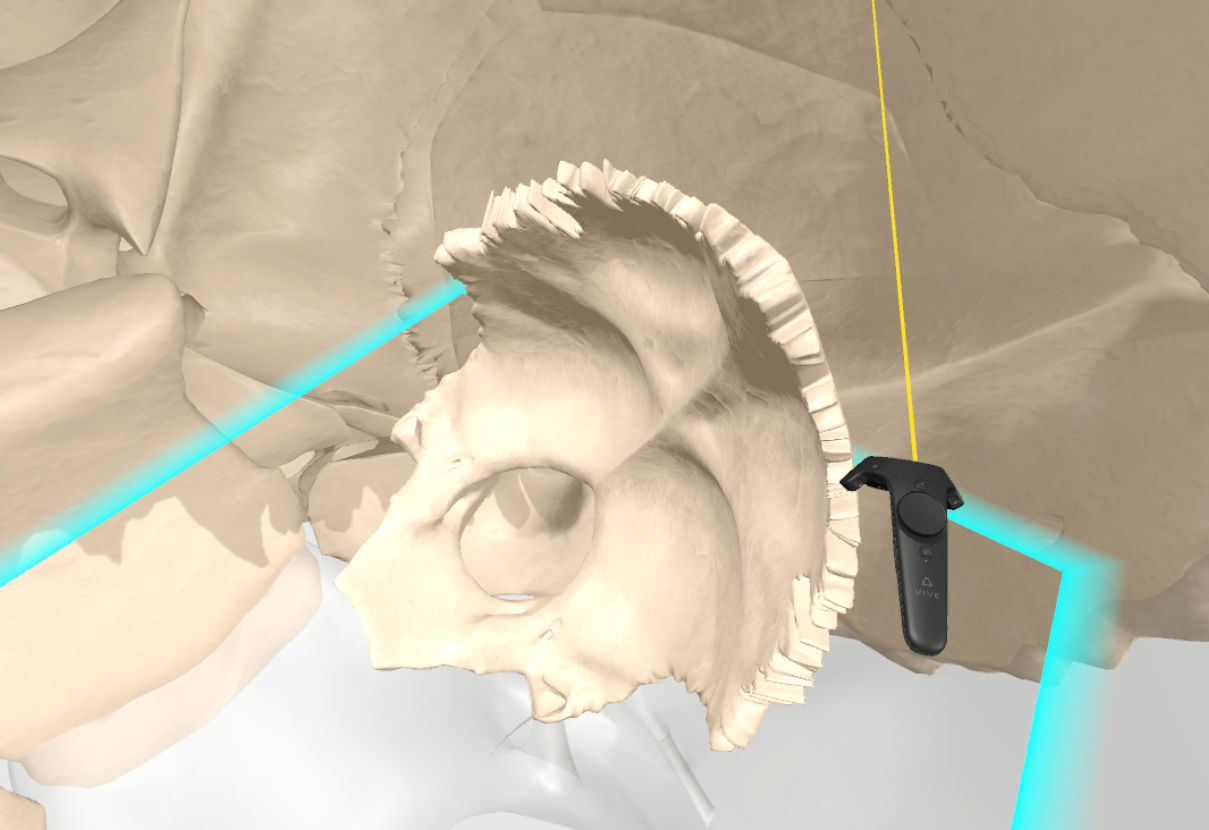}
	\caption{A smaller copy of a selected skull bones can be inspected closely. The original structure is shown semi-transparently in the background.}
	\label{fig:inspectObject}
\end{figure}

Due to the large scale of the skull base, the student may have difficulties to get a full image of single structures.
Some parts of a single bones may be occluded or hard to reach through navigation. 
The third page of the controller menu offers an \textit{Inspect Mode},
Here, the student can select single bones.
After selection, a ghost copy~\cite{Tan2001} of the bone is placed within the student's reach, while the original structure is displayed highlighted (see Fig.~\ref{fig:inspectObject}).
The student can use both controllers to grab and manipulate the structure directly with the virtual hand method~\cite{Robinett1992} which resembles real world interaction with an object.
The bone copy will move with the student when navigating to stay in reach all the time.
Even when the student leaves the inspection mode, the copy will stay present and can be manipulated. If the student wants to delete the copy or select another, she needs to switch back to the inspection mode menu page.
This parallel usage of inspecting a small copy while exploring the environment can help the student to achieve a better understanding of the 3D structures.

\subsection{Semi-immersive Teacher System}
According to the requirements for the teacher system, a balance between immersion and access to the real world should be provided. 
In order to be able to interact effectively within the virtual environment, a 3D input and output device is important for the teacher.
A semi-immersive system is a suitable compromise to benefit from immersive interaction and still being aware of the real world.

In our application, we use the \textit{zSpace} as a semi-immersive device that combines a stereoscopic display with head-tracking (so-called \textit{Fish Tank VR}~\cite{WAB93}).
Further, it provides a stylus as a 6 DOF pen-based input device, which is ideal for a natural input and especially suited for the required 3D sketch input.
This device was already used for anatomy education, i.e. the \textit{Visible Body} platform employed its benefit and Saalfeld et al.~\cite{Saalfeld_2016_VCBM} used it to sketch complex vascular structures.

The following sections present the user interface of the teacher system and discuss its functionality in detail.

\subsubsection{User Interface and Stylus Interaction}
\noindent\textbf{GUI.}
Following an analysis of 3D user interfaces for stereoscopic devices from Schild et al.~\cite{SBLM13}, a \textit{view fixed} interface is chosen. 
It is always rendered on top of the scene so it cannot be occluded by the 3D environment.

\begin{figure}[htb]
	\centering
	\includegraphics[width=0.715\linewidth]{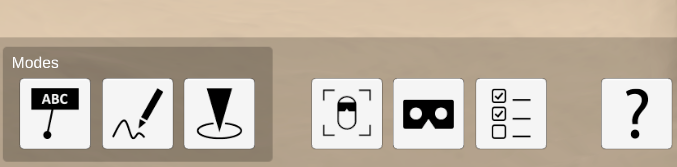}
	\includegraphics[width=0.27\linewidth]{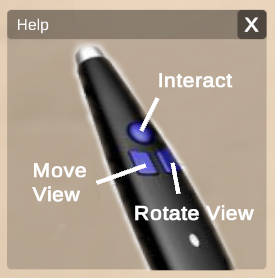}
	\caption{Left: the \textit{view fixed} user interface of the semi-immersive teacher system. From left to right: label mode, 3D sketch mode, landmark mode, center view on student, show student's view, show/hide annotations, help. Right: the help menu shows the button layout of the \textit{zSpace}'s stylus.
	}
	\label{fig:zspaceUI}
\end{figure}

Figure~\ref{fig:zspaceUI}, left, shows the menu bar attached to the bottom display frame of the \textit{zSpace}.
The menu bar is easily accessible with both hands to fit right-handed and left-handed users equally.

\noindent\textbf{Stylus Interaction.}
The stylus can be used for pointing tasks with ray casting, as the pen-like shape indicates the pointing direction intuitively.
Equally to the student's system, pointing at an object for selection is accompanied by highlighting of the object and vibro-tactile feedback of the stylus.

The stylus offers three input buttons, where the primary button controls different functionality based on the current mode.
The secondary and tertiary button enable the teacher to adjust the view at any time (in  Figure~\ref{fig:zspaceUI}, right, the help window with the button assignment is shown).
Usually the length of the laser beam is determined by the object that is hit. 
However, depending on the mode of interaction, the laser may have a predefined length, e.g., in the \textit{3D Sketch} mode.
\begin{figure}[htb]
	\centering
	
\end{figure}

\subsubsection{Guiding the Student's Navigation}
Following the motivation for this work, the main objective for guiding the student lies in supporting her navigation, in order to help her to follow a more productive path.
The teacher system offers two different approaches for this: indirectly with a \textit{landmark} or by \textit{manually reposition} the student.

\begin{figure}[htb]
	\centering
	\includegraphics[width=\linewidth]{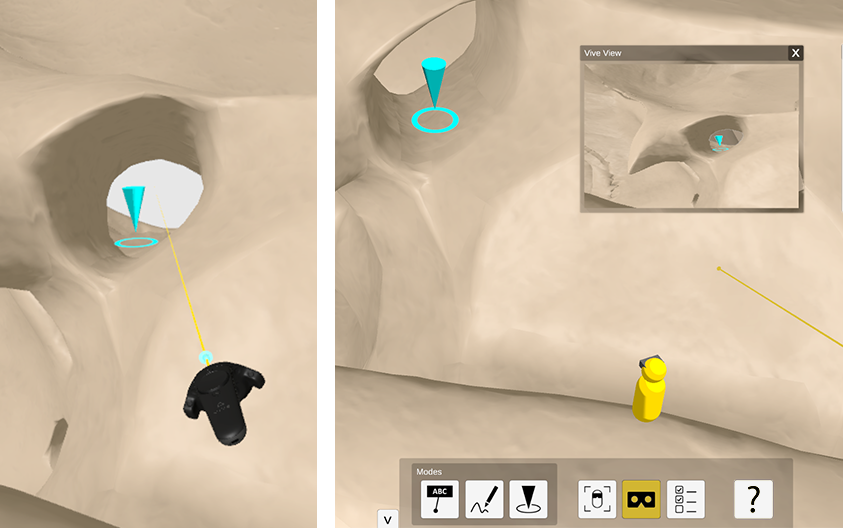}
	\caption{Left: The student's view on a landmark. Right: The corresponding view on the teacher system. The student's avatar can be seen, looking right at the landmark. The teacher can see the student's view with an extra window at the top right.}
	\label{fig:landmark}
\end{figure}

\noindent\textbf{Landmark.}
The teacher may indirectly lead the student to a desired location, through placing a landmark at the target position.
The landmark is represented by a simple cyan colored cel-shaded cone that stands out within the environment (see Fig.~\ref{fig:landmark}). An animation of the cone jumping up and down draws additional attention towards the landmark.

To place a landmark into the environment, the teacher needs to activate the \textit{Landmark Interaction Mode}. As interesting positions usually lie on a surface of a structure, the landmark position can be set via ray casting onto the 3D model of the skull base.
After a landmark is placed, it is synchronized over the network and the student is informed through a vibration feedback of her controller. Additionally, an arrow is placed at the student's  controller facing into the direction of the landmark (see Fig. \ref{fig:landmark}).
The student still has to navigate on her own, but wayfinding will be improved as she now has an object of orientation.

\noindent\textbf{Reposition Student.}
Even though the landmark can lead the student to a desired location, it can still be helpful to actively change the students' position.
If the teacher wants to start the lesson at a specific location, she can directly place the student as desired.

To reposition the student, the teacher can \textit{pick up} the student's avatar with the stylus and releasing it somewhere else. Only the target position is synchronized to the students' systems to prevent cybersickness by getting moved around.

\subsubsection{Creating Labels}
\begin{figure}[htb]
	\centering
	\includegraphics[width=0.8\linewidth]{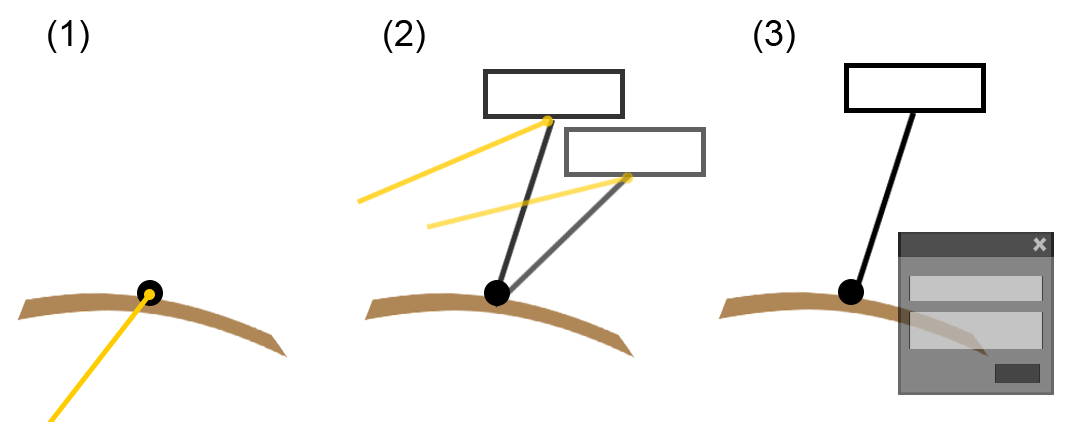}
	\caption{Labels can be created by (1) ray casting onto the surface, (2) dragging the label to the desired location and (3) edit its content within an extra window.}
	\label{fig:label_steps}
\end{figure}

\begin{figure}[htb]
	\centering
	\includegraphics[width=0.75\linewidth]{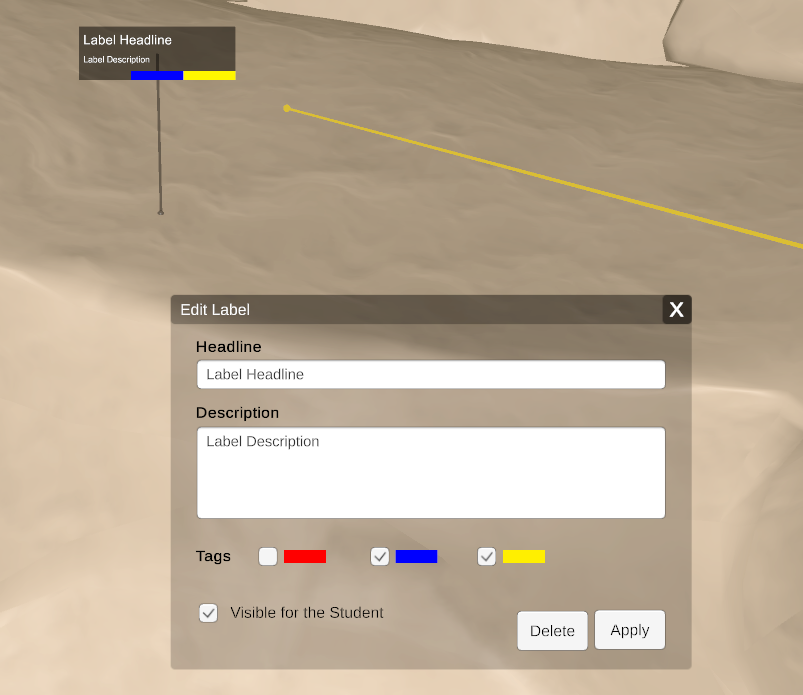}
	\caption{A label that is placed in the environment can be edited via the label window.}
	\label{fig:label_window}
\end{figure}

The teacher should be able to create labels in the virtual scene and edit them while tutoring.
We follow on Pick et al.~\cite{Pick2016} who present a workflow to create, review and modify labels in VR. 
Labels can only be positioned on a surface, in order to assure a direct link between object and information.
The creation of a label is realized in three steps (see Fig.~\ref{fig:label_steps}).
After ray casting onto the desired position of the label starting point, the teacher presses the primary button, creates a label and can drag it to the desired location in the 3D space.
Labels are always facing the student, thus, its orientation does not have to be defined by the teacher.
The finished label placement results in a dialogue being displayed, for editing the label contents.
Here, the teacher can enter a headline, a description text and a tag in form of a color (red, blue yellow). This tag is useful to hint the student if the labels is related to e.g. arteries, veins and muscles passing through foramina (see Fig.~\ref{fig:label_window}).

\subsubsection{Creating 3D Sketched Annotations}
A dedicated input technique is provided for the teacher to allow her to draw structures such as vessels and nerves freely in 3D space to create illustrations for explanation and highlight certain areas.

\begin{figure}[htb]
	\centering
	\includegraphics[width=\linewidth]{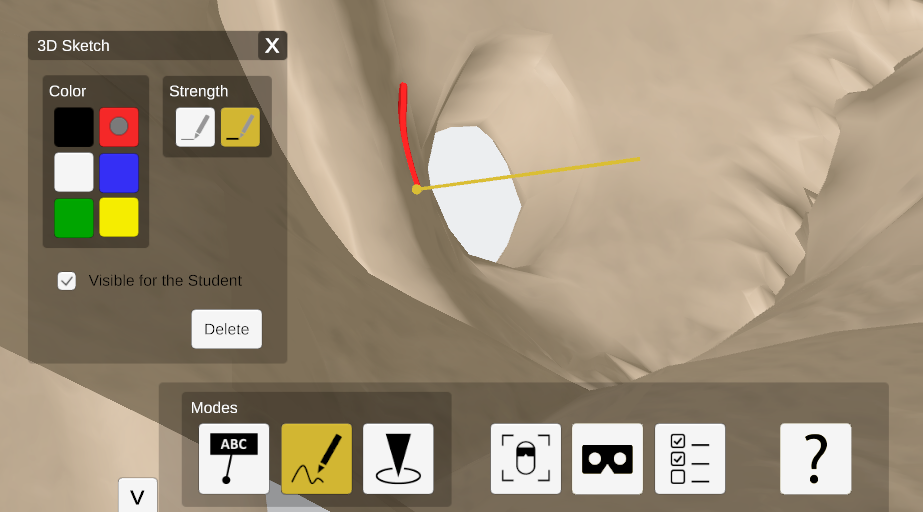}
	\caption{The \textit{3D Sketch Mode} is active. The window comprising sketch settings is opened while a red stroke is sketched directly onto the surface.}
	\label{fig:zspace_sketch}
\end{figure}

The \textit{3D Sketch Interaction Mode} opens a window with settings for sketching. The teacher can choose from a color palette and two different brush sizes (see Fig.~\ref{fig:zspace_sketch}).
Furthermore, the laser beam is set to a predefined length to allow the teacher to sketch in mid-air. If the beam hits a surface, it is shortened to lie on the surface.
During pressing the primary button, the sketch is created at the laser beam tip and continuously synchronized with the student system.
The teacher draws a centerline which is enclosed by a 3D tube during sketching.
This allows to apply a cel-shading onto the created tube surface.


\subsubsection{Hiding Annotations}
In order to structure a tutoring lesson, the teacher may not want to display all created labels or sketches at the same time.
Deleting those not needed is impractical, as they might be needed again later.
Therefore, the teacher has the possibility to hide annotations for the student and display them again when needed.
However, it is currently not possible to save labels or annotations for later use.

\subsubsection{Adjusting the View and Observing the Student}
The teacher needs to navigate within the virtual environment and in order to focus on areas of interest and to observe the student.
The base of the skull is a hollow space.
While the student's fully immersive system is suitable for this environment, the \textit{zSpace} usually constrains the view to one direction.
To solve this problem, the teacher is able to translate and rotate its own view with the stylus.
While pressing the left button of the stylus, the teacher's view can be translated on all three axes according to the stylus movement.
While pressing the right button, the teacher's view is rotated around itself. The vertical rotation is constrained to a maximum angle to prevent the teacher from having an upside down view.

If the teacher loses track of the student, which can happen easily if the student teleports, the menu bar offers a button to focus on the student.
After pressing it, the teachers view is translated so that she focuses on the student's avatar.

Additionally, the teacher can show the \textit{Student View Window}, which shows the viewport of student.
Instead of transmitting a video stream, only the student's head position is synchronized to the teacher. This position is attached to a second virtual camera on the teacher's system. The grabbed image of this camera is shown in the Student View Window.

\section{Technical Details}
This section gives more technical details on the used VR devices and the realization of the application.
The application was developed with the Unity game engine

\begin{figure}[htb]
	\centering
	\includegraphics[width=0.3\linewidth]{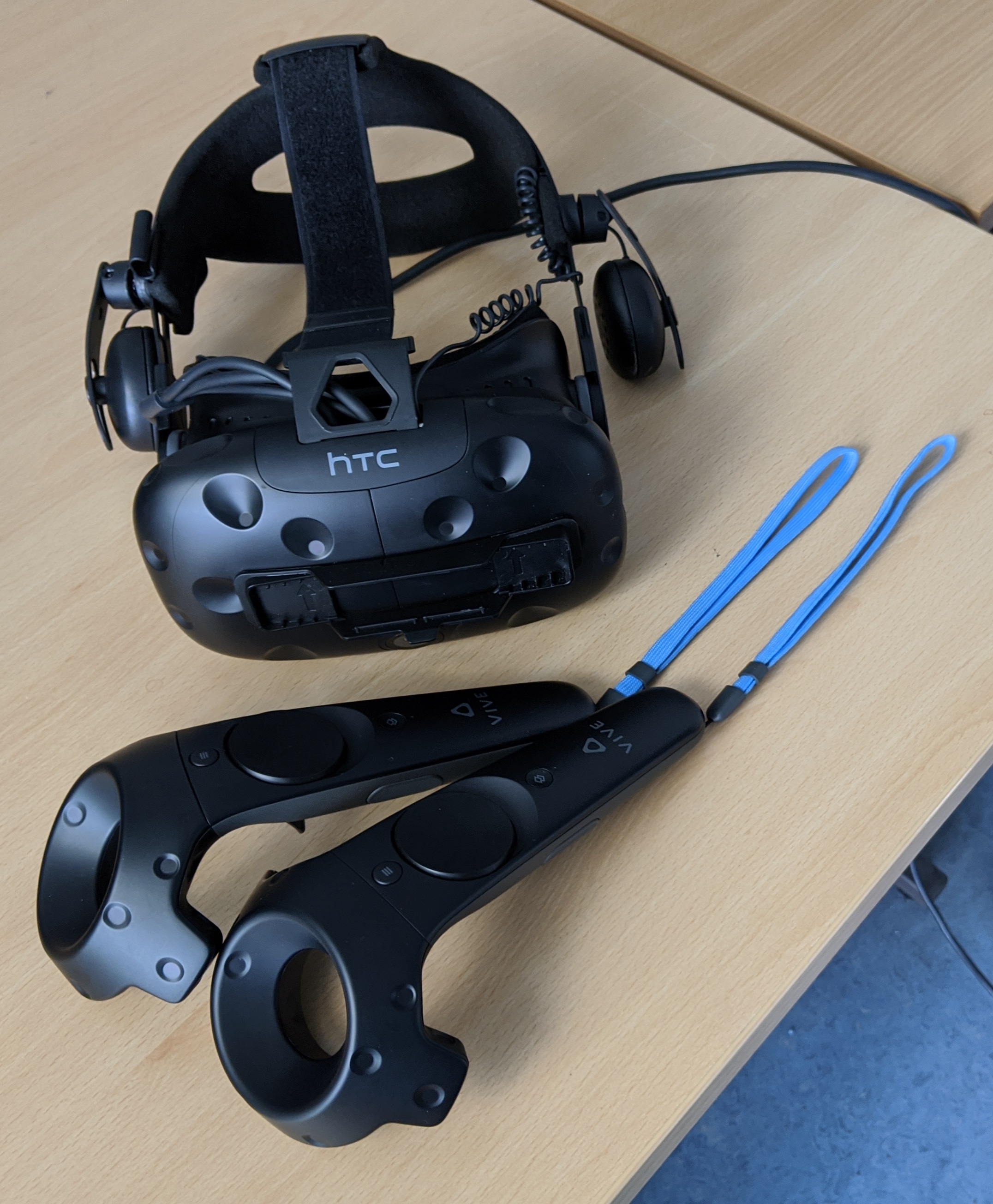}
	\hspace{0.5cm}
	\includegraphics[width=0.405\linewidth]{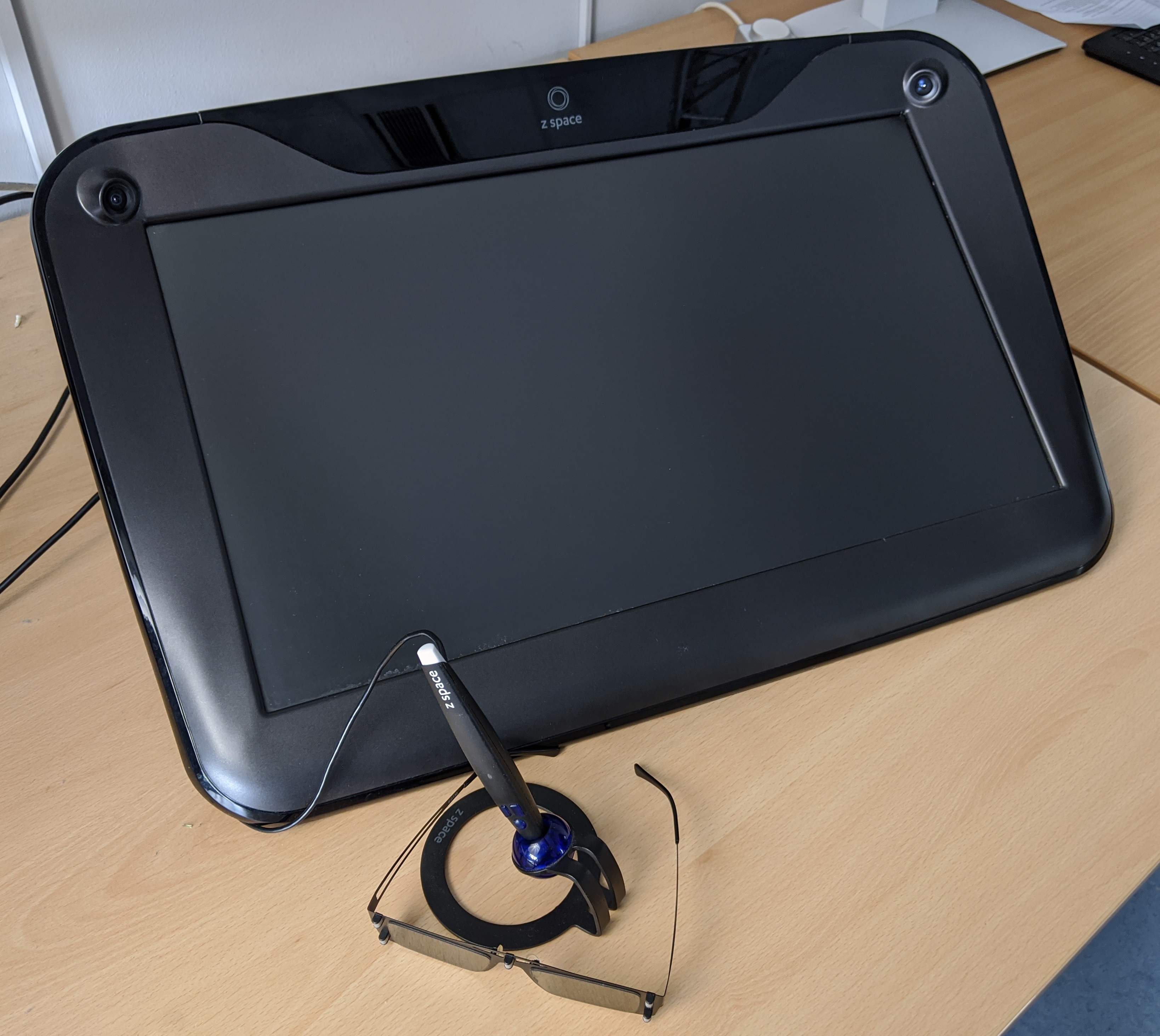}
	\caption{Left: the \textit{HTC Vive} system. Right: the \textit{zSpace} system.}
	\label{fig:hardware}
\end{figure}

\noindent\textbf{Fully Immersive HTC Vive.}
The standard setup of the \textit{HTC Vive} comprises the headset, two controllers and two \textit{lighthouse} tracking stations (see Fig.~\ref{fig:hardware}).
Each of the two OLED displays of the headset have a resolution of 1080 $\times$ 1200 pixels with a refresh rate of 90\,Hz.
The headset offers a \textit{field of view} of approximately 110\textdegree{} to achieve full immersion.
The tracking is achieved through various sensors, such as gyroscope or accelerometer and two \textit{lighthouse} base stations that emit pulses of infrared light.
This roughly allows a 5\,m $\times$ 5\,m area of tracked movement.
The \textit{HTC Vive} is supported in the Unity game engine through the \textit{SteamVR} plugin.

\noindent\textbf{Semi-Immersive zSpace.}
The \textit{zSpace} is a stereoscopic 3D monitor combined with headtracking and a stylus input device (see Fig.~\ref{fig:hardware}).
The user wears passive polarized 3D glasses that allow for the stereoscopic view.
Headtracking is achieved optically via tracking the worn glasses.
Furthermore, a tracked stylus is provided that allows 6 DOF pen-based input.
The zSpace can be integrated into Unity with the \textit{zCore} plugin.

\noindent\textbf{Networking.}
As both VR systems are operating on different computers, these need to be connected via network.
We realized this with Unity's own directly integrated multiplayer network functionality \textit{UNET}.
In UNET, there is no dedicated server, instead, one client is taking the role of the host as well.
In our application, this is done by the teacher system, while the student system connects to this system as a remote client.

\section{Evaluation}
For both the student and teacher system, a qualitative user study was performed to give first impressions on usability issues and how the one-on-one tutoring approach helps to improve the VR experience of the student.
After completing a series of tasks, the users are presented a questionnaire that addresses the requirements accumulated in Section~\ref{sec:requirements}, questions regarding the usability, felt immersion and presence as well as how they evaluate the different functionalities.
Additionally, the users are asked to think aloud.

\subsection{Participants}
Both systems were tested with six participants (5 males, 1 female) between the age of 22 and 31.
Three of them had previously experienced VR applications.
The other three were familiar with 3D input devices through game consoles such as the \textit{Nintendo Wii}.
None of the participants had a medical background.
Thus, an evaluation with the target audience is still necessary to focus on the applicability for anatomy learning.

\subsection{Procedure}
Each participant used the student system first, filled out a questionnaire, used the teacher system second, and filled out another questionnaire.
The whole study took between 40 -- 70 minutes, depending on the speed of the participant.
In the following, the individual steps are described in more detail.

\subsubsection{Student system}
Each run started by asking each participant regarding demographic data as well as previous experience with VR, 3D input devices and 3D medical data.

The role of the teacher is taken by the instructor.
First, the user is introduced to the navigation concepts of \textit{free flying} and the \textit{teleport} inside the enlarged human skull.
It is explained that she can be positioned to a desired location by the instructor.
This procedure is tested once.
Following, the concept of accessing detailed information on structures by pointing at them is explained.
Finally, the participant is introduced to the \textit{inspection mode}.
She can try out these functionalities individually and is asked to inform the instructor as soon as she feels sufficiently familiarized with the controls.

After the familiarization phase, the user is asked to visit three foramina of the base of skull (\textit{Canalis opticus}, \textit{Left Foramen Ovale}, \textit{Right Foramen Spinosum}).
This task was chosen by an experienced anatomy teacher based on its relevance.
They vary in size and their location affords navigation between them. The position of each foramina is indicated via landmarks, resulting in an indication arrow on the participant's controller.
The participants are free to decide which navigation method to use.
As soon as the user arrives at the target location, the foramina is circled with the 3D sketch input by the instructor.
The participant is introduced to the name of the structure while creating a label to annotate the foramina.
This procedure is repeated for the other two structures.

After the third foramina structure has been examined, the participant is asked to navigate back to one of the previously shown structures.

\subsubsection{Teacher system}
The user study of the teacher system followed afterwards.
Hereby, the participants take advantage of the fact that they are already familiar with the basic functionality by having experienced it through the student system.

The role of the student is taken by the instructor that simply moves around the environment spontaneously.
The state of the student system is only partly relevant for the usability evaluation of the teacher system.
It suffices to have a moving student avatar to follow.
Each participant has to solve a sequence of the following interaction tasks of the teacher system.
The participant should:
\begin{itemize}
    \item open the \textit{Vive View} Window and drag it from the top right position to the top left corner to learn how to handle windows,
    \item translate and rotate the view to focus a foramen to train view adjustment,
    \item create two landmarks at different locations,
    \item use the 3D sketching tool to circle the focused foramina and adjust its color,    
    \item select and delete the sketch with the \textit{3D sketch} window, and
    \item create a label near the focused foramina with an arbitrary text.
\end{itemize}

\subsection{Results}
\subsubsection{Student System}
In the following, the results regarding presence, immersion and usability of the different interaction methods are presented.
The results on navigation are presented separately in the next section.

\noindent\textbf{Immersion.}
The participants generally stated to have a strong sense of immersion and evaluated the interaction as very natural.
Especially the position tracking which allows physical movement to be mapped directly to the virtual world was considered to highly contribute to a real-world feeling.
One participant was even frightened at the first moment he looked down to the inner skull surface from his elevated position within the skull.
Especially the live synchronization of the sketches fascinated the users.
One participant remarked, regarding the live sketching, that he is now realizing, there is really another person involved.

\noindent\textbf{Presentation of Text and Structures.}
The presentation of textual information was mainly evaluated positively.
Every participant stated to easily understand what structures the information referred to.
This comprised the \textit{handbook} functionality, as well as the labels.
Only one participant stated that she had problems reading texts, as she did not manage to adjust the HMD to achieve a sharp view.

The scaled skull model allowed the participants to have a detailed look on structures.
However, the majority stated that they do not feel able to project the acquired spatial knowledge to a smaller real-size model.
Being able to recognize structures within the scaled model does not necessarily imply that the knowledge can be transferred to the smaller real skull.
One of the participants, however, indicated that the 2D illustrations of hovered structures displayed in the handbook were very helpful.

\noindent\textbf{Inspection of Single Structures.}
Only one participant used the inspection of a single structure again after the familiarization phase.
A reason may be that the participants did not have a real interest in the specific structures, due to the missing medical background.
The clear focus on the navigation tasks may be another reason.
The one person using the functionality was especially fascinated by taking the object along with him while continuing to navigate.

\noindent\textbf{Easy to Learn Interface.}
The majority of the participants evaluated the displayed labels that were attached to the controllers as rather helpful.
None of the participants displayed this help after the familiarization phase.
As they were still able to perform the demanded navigation tasks, the interface seems to be easily learnable.
The questionnaire further confirmed this, as none of the students stated to be confused about the interaction.

\noindent\textbf{Navigation, Orientation and Cybersickness.}
All of the participants mainly used the \textit{free flying} navigation.
Even though it was assumed that indicating the steering direction through the orientation of the controller might not be intuitive, every one of the participants rated this method intuitive.

The \textit{teleport} technique, however, was rated as hard to access by half of the users.
They generally did not like having to activate a specific interaction mode to access the teleport.
Despite the large scale of the environment, a majority found the free navigation rather sufficient for the navigation tasks.
While none of the participants felt disoriented while freely navigating, each of them evaluated the teleport to cause at least a slight disorientation.

An even higher rate of disorientation was linked to the process of the teacher altering the position of the user on the student system.
This function was tested once in the familiarization phase and was announced beforehand.
A common reaction was, that the participants immediately tried to navigate back to the initial position, in order to orient themselves.
The evaluation clearly shows that a direct cut to a new location has negative effects on the orientation.
The evaluated navigation methods did not evoke cybersickness in any of the participants.

\noindent\textbf{Wayfinding.}
The landmark as a hint for wayfinding was highly acclaimed by all participants and could be successfully located by each one.
Notable, however, one participant intentionally flew directly through walls to get to the landmark quicker.

All but one participant were able to navigate to a previously visited location without a landmark. However, some stated that they felt unsure about being successful until the instructor confirmed the correct location.
Some of the participants navigated to a position that gives a broader overview over the environment in order to look for the destination.

\subsubsection{Teacher System}
All participants found it easy to understand the interaction technique that the stylus presents.
This is not surprising, as it combines natural pen-input with the previously experienced ray-casting method of the fully immersive system.
The users found the handling of the windows comprehensible as the concept is known through the common \textit{WIMP} approach.
They mainly enjoyed being able to reposition the windows individually.

The view adjustment was the big downside of the teacher system.
None of the participants felt confident when translating or rotating the viewport.
They felt it was cumbersome to use.
This problem mainly arises due to the required 360\textdegree{} view, a scenario that is sub optimal for the \textit{zSpace} system.

The 3D sketching was the most positively acclaimed interaction of the teacher system.
The majority of participants stated they would like to use this interaction technique frequently.
Obviously, several participants enjoyed this feature, as two of them asked to try this feature again outside of the evaluation process.
Half of the participants felt very confident using the system and one of them pointed out that he likes the mechanism to be able to directly draw onto surfaces.
However, as the general hand position to use the stylus does not necessarily allow the user to support her arm with the table, inaccuracies seemed to increase with time.

The creation of labels came out neutral in the evaluation.
While two participants found the positioning of the labels easy, three users found the technique too complex.
Three participants felt that the text input through the keyboard was cumbersome, as the user has to switch between two input devices.
One user, on the other hand, strongly appreciated the possibility for text input.

\subsection{Discussion}
The results for the student system were mainly positive.
The users felt immersed in the virtual world and evaluated the interaction techniques as natural and intuitive.
Especially the landmark turned out to be a great tool to guide the student to a specific location.
The landmark concept might be an interesting field of investigation, in order to evaluate if it hinders the user to build up a cognitive map of the virtual environment, as it makes wayfinding very easy.

Regarding the different navigation techniques, the teleport method was not able to convince the users.
Furthermore, the immediate change of positions that was caused by the teacher was confusing for the participants.
Here, a transition instead of a direct cut could improve the experience.
Additionally, users of the student system wished for more input possibilities, e.g., being able to point and mark specific locations within the environment.

Interestingly, the semi-immersive \textit{zSpace} had problems to fulfill some of the requirements for the teacher system.
The device usually unfolds its potential when only one object of attention is in the center of the scene.
In our application scenario, however, the focus of attention is not on a single view direction, but navigating the view in a 360\textdegree{} dome, i.e. the skull base.
Therefore, different view manipulation techniques should be investigated.
The \textit{zSpace} as a \textit{Fish Tank VR} system does not provide a natural input to achieve the needed 360\textdegree{} view.

Despite the downsides of the semi-immersive system, the evaluation showed that navigation tasks can be influenced positively through guiding a student in a shared virtual environment.
Concepts such as the landmark integration, might help to make VR learning more effective.

\section{Summary \& Conclusion}
We introduced an approach for a VR one-on-one tutoring system to support anatomy education.
Current VR technology offers new dimensions of exploring anatomy structures, but usually only provides a single user experience.
We presented a shared virtual environment where guidance from a teacher is provided.
We put focus on the improvement of the student's navigation and the creation of annotations within the virtual world.
An asymmetric setting was realized, using the semi-immersive \textit{zSpace} and the fully immersive \textit{HTC Vive} that share a network connection. 

A qualitative user study was conducted, where six participants used the system in both roles consecutively. 
The participants assessed the fully immersive student system mainly positive, stating that they felt immersed and that the interaction techniques are natural.
The teleportation and instant teleport triggered by the teacher needs to be improved.
The semi-immersive teacher system was also mainly rated positively.
Here, already known WIMP concepts made it easier to use the system. However, navigation inside the hollow dome of the skull was challenging for the participants and needs to be improved.

Even though one-on-one tutoring is an effective tool of instruction, it is not always applicable, due to a high number of educators needed.
Therefore, possibilities to extend the presented approach to a classroom setting, where the teacher guides several students, is an interesting topic for future research.
Another way to reach more students in one teaching session is by integrating a possibility for remote collaboration, where the students can be at different physical locations.
In order to evaluate the applicability of the system in the actual education context, a future formal analysis with the target audience (students and teaching staff involved in a medical curriculum) is necessary.
The evaluation of navigation techniques should include an assessment how well they support or prevent building up a  mental map of the environment, which could be an important indicator for the learning outcome.


\bibliographystyle{unsrt}  
\bibliography{vrAnatomyShared}

\newcommand{\etalchar}[1]{$^{#1}$}
\begin{thebibliography}{\uppercase{SBLJM13}}

\bibitem[BHB{\etalchar{*}}07]{BHBSPFD07}
\textsc{Brenton H., Hernandez J., Bello F., Strutton P., Purkayastha S., Firth
  T., Darzi A.}:
\newblock Using multimedia and web3d to enhance anatomy teaching.
\newblock \emph{Computers \& Education 49}, 1 (2007), 32--53.

\bibitem[Blo84]{Blo84}
\textsc{Bloom B.~S.}:
\newblock The 2 sigma problem: The search for methods of group instruction as
  effective as one-to-one tutoring.
\newblock \emph{Educational researcher 13}, 6 (1984), 4--16.

\bibitem[BMS{\etalchar{*}}13]{Birr:2013}
\textsc{Birr S., M{\"{o}}nch J., Sommerfeld D., Preim U., Preim B.}:
\newblock The liveranatomyexplorer: {A} webgl-based surgical teaching tool.
\newblock \emph{{IEEE} Computer Graphics and Applications 33}, 5 (2013),
  48--58.

\bibitem[BPT01]{BPT01}
\textsc{Bouras C., Philopoulos A., Tsiatsos T.}:
\newblock e-learning through distributed virtual environments.
\newblock \emph{Journal of Network and Computer Applications 24}, 3 (2001),
  175--199.

\bibitem[Bri90]{Bri90}
\textsc{Bricken W.}:
\newblock \emph{Learning in Virtual Reality}.
\newblock Tech. rep., University of Washington, 1990.

\bibitem[Bro00]{Bro00}
\textsc{Brown J.~R.}:
\newblock Enabling educational collaboration—a new shared reality.
\newblock \emph{Computers \& Graphics 24}, 2 (2000), 289--292.

\bibitem[Cun18]{Cun18}
\textsc{Cunningham D.~J.}:
\newblock \emph{Cunningham's textbook of anatomy}.
\newblock W. Wood, 1818.

\bibitem[DHB01]{DHB01}
\textsc{Dev P., Hoffer E.~P., Barnett G.~O.}:
\newblock Computers in medical education.
\newblock In \emph{Medical informatics}. Springer, 2001, pp.~610--637.

\bibitem[EM06]{EM06}
\textsc{Evens M., Michael J.}:
\newblock \emph{One-on-one tutoring by humans and computers}.
\newblock Psychology Press, 2006.

\bibitem[Ham82]{Ham82}
\textsc{Hamilton W.~J.}:
\newblock \emph{Textbook of human anatomy}.
\newblock Springer, 1982.

\bibitem[HPP{\etalchar{*}}96]{HPPRSST96}
\textsc{Hohne K.-H., Pflesser B., Pommert A., Riemer M., Schiemann T., Schubert
  R., Tiede U.}:
\newblock A'virtual body'model for surgical education and rehearsal.
\newblock \emph{Computer 29}, 1 (1996), 25--31.

\bibitem[HRL10]{HRL10}
\textsc{Huang H.-M., Rauch U., Liaw S.-S.}:
\newblock Investigating learners’ attitudes toward virtual reality learning
  environments: Based on a constructivist approach.
\newblock \emph{Computers \& Education 55}, 3 (2010), 1171--1182.

\bibitem[HS08]{HS08}
\textsc{Hanson K., Shelton B.~E.}:
\newblock Design and development of virtual reality: Analysis of challenges
  faced by educators.
\newblock \emph{Educational Technology \& Society 11}, 1 (2008), 118--131.

\bibitem[HV97]{HV97}
\textsc{Hoffman H., Vu D.}:
\newblock Virtual reality: teaching tool of the twenty-first century?.
\newblock \emph{Academic Medicine 72}, 12 (1997), 1076--81.

\bibitem[JF00]{JF00}
\textsc{Jackson R.~L., Fagan E.}:
\newblock Collaboration and learning within immersive virtual reality.
\newblock In \emph{Proc. of Collaborative virtual environments} (2000), ACM,
  pp.~83--92.

\bibitem[JRL{\etalchar{*}}98]{JRLVBM98}
\textsc{Johnson A., Roussos M., Leigh J., Vasilakis C., Barnes C., Moher T.}:
\newblock The nice project: Learning together in a virtual world.
\newblock In \emph{Proc. of Virtual Reality Annual International Symposium,}
  (1998), IEEE, pp.~176--183.

\bibitem[JTW99]{JTW99}
\textsc{Jackson R.~L., Taylor W., Winn W.}:
\newblock Peer collaboration and virtual environments: a preliminary
  investigation of multi-participant virtual reality applied in science
  education.
\newblock In \emph{Proc. of ACM symposium on Applied computing} (1999),
  pp.~121--125.

\bibitem[KSJ{\etalchar{*}}13]{kraima2013toward}
\textsc{Kraima A.~C., Smit N.~N., Jansma D., Wallner C., Bleys R., Velde C.
  v.~d., Botha C., DeRuiter M.~C.}:
\newblock Toward a highly-detailed 3d pelvic model: Approaching an
  ultra-specific level for surgical simulation and anatomical education.
\newblock \emph{Clinical Anatomy 26}, 3 (2013), 333--338.

\bibitem[LC00]{LC00}
\textsc{Lander J., CONTENT G.}:
\newblock Shades of disney: Opaquing a 3d world.
\newblock \emph{Game Developer Magazine 7}, 3 (2000), 15--20.

\bibitem[LCDL{\etalchar{*}}16]{LDLGRA16}
\textsc{Le~Ch{\'e}n{\'e}chal M., Duval T., Lacoche J., Gouranton V., Royan J.,
  Arnaldi B.}:
\newblock When the giant meets the ant an asymmetric approach for collaborative
  object manipulation.
\newblock In \emph{Proc. of IEEE 3D User Interfaces} (2016), pp.~277--278.

\bibitem[MJOG99]{MJOG99}
\textsc{Moher T., Johnson A., Ohlsson S., Gillingham M.}:
\newblock Bridging strategies for vr-based learning.
\newblock In \emph{Proc. of ACM SIGCHI conference on Human Factors in Computing
  Systems} (1999), pp.~536--543.

\bibitem[MOA{\etalchar{*}}19]{Maresky:2019}
\textsc{Maresky H., Oikonomou A., Ali I., Ditkofsky N., Pakkal M., Ballyk B.}:
\newblock {Virtual reality and cardiac anatomy: Exploring immersive
  three-dimensional cardiac imaging, a pilot study in undergraduate medical
  anatomy education}.
\newblock \emph{Clinical Anatomy 32}, 2 (2019), 238--243.

\bibitem[MWS17]{Marks:2017}
\textsc{Marks S., White D., Singh M.}:
\newblock Getting up your nose: a virtual reality education tool for nasal
  cavity anatomy.
\newblock In \emph{Proc. of SIGGRAPH Asia 2017 symposium on education} (2017),
  pp.~1--7.

\bibitem[PP00]{PP00}
\textsc{Putz R., Pabst R.}:
\newblock \emph{Sobotta, Atlas of Human Anatomy Volume 1 - Head, Neck, Upper
  Limb}.
\newblock Urban \& Fischer, 2000.

\bibitem[PPS99]{PPS99}
\textsc{Pitt I., Preim B., Schlechtweg S.}:
\newblock An evaluation of interaction techniques for the exploration of
  3d-illustrations.
\newblock In \emph{Software-Ergonomie’99}. Springer, 1999, pp.~275--286.

\bibitem[PPS19]{Pohlandt:2019}
\textsc{Pohlandt D., Preim B., Saalfeld P.}:
\newblock {Supporting Anatomy Education with a 3D Puzzle in a Virtual Reality
  Environment}.
\newblock In \emph{Mensch und Computer} (2019).

\bibitem[PRS97]{PRS97}
\textsc{Preim B., Raab A., Strothotte T.}:
\newblock Coherent zooming of illustrations with 3d-graphics and text.
\newblock In \emph{Graphics Interface} (1997), vol.~97, pp.~105--113.

\bibitem[PS18]{Preim:2018}
\textsc{Preim B., Saalfeld P.}:
\newblock {A Survey of Virtual Human Anatomy Education Systems}.
\newblock \emph{Computers \& Graphics 71} (2018), 132--53.

\bibitem[Pso95]{Pso95}
\textsc{Psotka J.}:
\newblock Immersive training systems: Virtual reality and education and
  training.
\newblock \emph{Instructional science 23}, 5 (1995), 405--431.

\bibitem[PWHK16]{Pick2016}
\textsc{{Pick} S., {Weyers} B., {Hentschel} B., {Kuhlen} T.~W.}:
\newblock Design and evaluation of data annotation workflows for cave-like
  virtual environments.
\newblock \emph{IEEE Transactions on Visualization and Computer Graphics 22}, 4
  (2016), 1452--1461.

\bibitem[RH92]{Robinett1992}
\textsc{Robinett W., Holloway R.}:
\newblock Implementation of flying, scaling and grabbing in virtual worlds.
\newblock In \emph{Proc. of ACM Symposium on Interactive 3D Graphics} (1992),
  pp.~189--–192.

\bibitem[RPDS00]{RPDS00}
\textsc{Ritter F., Preim B., Deussen O., Strothotte T.}:
\newblock Using a 3d puzzle as a metaphor for learning spatial relations.
\newblock In \emph{Proc. of Graphics Interface} (2000), pp.~171--178.

\bibitem[RSM92]{RSM92}
\textsc{Regian J., Shebilske W.~L., Monk J.~M.}:
\newblock Virtual reality: an instructional medium for visual-spatial tasks.
\newblock \emph{Journal of Communication 42}, 4 (1992), 136--149.

\bibitem[SAK10]{SAK10}
\textsc{Sugand K., Abrahams P., Khurana A.}:
\newblock The anatomy of anatomy: a review for its modernization.
\newblock \emph{Anatomical sciences education 3}, 2 (2010), 83--93.

\bibitem[SBLJM13]{SBLM13}
\textsc{Schild J., B{\"o}licke L., LaViola~Jr J.~J., Masuch M.}:
\newblock Creating and analyzing stereoscopic 3d graphical user interfaces in
  digital games.
\newblock In \emph{Proc. of the ACM SIGCHI conference on human factors in
  computing systems} (2013), pp.~169--178.

\bibitem[Sch17]{Schmeier2017}
\textsc{Schmeier A.}:
\newblock \emph{Student and Teacher Meet in a Shared Virtual Environment: A VR
  One-on-One Tutoring System to Support Anatomy Education}.
\newblock Master's thesis, Otto-von-Guericke University Magdeburg, Germany,
  Dept. of Computer Science, 2017.

\bibitem[SSPOJ16]{Saalfeld_2016_VCBM}
\textsc{Saalfeld P., Stojnic A., Preim B., Oeltze-Jafra S.}:
\newblock {Semi-Immersive 3D Sketching of Vascular Structures for Medical
  Education}.
\newblock In \emph{{Proc. of Eurographics Workshop on Visual Computing for
  Biology and Medicine (EG VCBM)}} (2016), pp.~123--132.

\bibitem[TN02]{TN02}
\textsc{Tax{\'e}n G., Naeve A.}:
\newblock A system for exploring open issues in vr-based education.
\newblock \emph{Computers \& Graphics 26}, 4 (2002), 593--598.

\bibitem[TRC01]{Tan2001}
\textsc{Tan D.~S., Robertson G.~G., Czerwinski M.}:
\newblock Exploring 3d navigation: Combining speed-coupled flying with
  orbiting.
\newblock In \emph{Proc. of ACM SIGCHI Conference on Human Factors in Computing
  Systems} (2001), pp.~418–--425.

\bibitem[WAB93]{WAB93}
\textsc{Ware C., Arthur K., Booth K.~S.}:
\newblock Fish tank virtual reality.
\newblock In \emph{Proc. of INTERACT and CHI conference on Human factors in
  computing systems} (1993), pp.~37--42.

\end{thebibliography}



\end{document}